\documentclass[letterpaper,twocolumn,english,nopreprint,prd,nofootinbib,preprintnumbers,showpacs,showkeys,floatfix]{revtex4}
\usepackage[T1]{fontenc}
\usepackage[latin1]{inputenc}
\usepackage{graphicx}
\usepackage{esint}

\makeatletter

\providecommand{\tabularnewline}{\\}

\@ifundefined{textcolor}{}
{%
 \definecolor{BLACK}{gray}{0}
 \definecolor{WHITE}{gray}{1}
 \definecolor{RED}{rgb}{1,0,0}
 \definecolor{GREEN}{rgb}{0,1,0}
 \definecolor{BLUE}{rgb}{0,0,1}
 \definecolor{CYAN}{cmyk}{1,0,0,0}
 \definecolor{MAGENTA}{cmyk}{0,1,0,0}
 \definecolor{YELLOW}{cmyk}{0,0,1,0}
 }

\def\Log{{\rm Log}}\def\msbar{\overline{MS}}

\makeatother

\usepackage{babel}

\makeatother

\usepackage{babel}

\begin{document}

\preprint{arXiv:0812.357}

\title{Regularization, Renormalization, and Dimensional Analysis: \\
 \textmd{\textit{Dimensional Regularization Meets Freshman E\&M}}\footnote{%
Published in Am.J.Phys. 79 (2011) 306 
}
}

\author{Fredrick~Olness \& Randall Scalise}

\affiliation{Department of Physics, Southern Methodist University, Dallas, TX
75275-0175, U.S.A. }
\begin{abstract}
\noindent We illustrate the dimensional regularization (DR) technique
using a simple problem from elementary electrostatics. This example
illustrates the virtues of DR without the complications of a full
quantum field theory calculation. We contrast the DR approach with
the cutoff regularization approach, and demonstrate that DR preserves
the translational symmetry. We then introduce a Minimal Subtraction
($MS$) and a Modified Minimal Subtraction ($\overline{MS}$) scheme
to renormalize the result. Finally, we consider dimensional transmutation
as encountered in the case of compact extra-dimensions.

\vspace{20pt}

\end{abstract}


\keywords{Renormalization, Dimensional Regularization, Regularization, Gauge
Symmetries 
\vspace{1.4in}}

\date{\today}
\maketitle

\tableofcontents{}

\vfill
\newpage

\section{Dimensional Regularization }

\subsection{Introduction and Motivation}

In 1999, Gerardus 't Hooft and Martinus J.G.\,Veltman received the
Nobel Prize in Physics\citep{physicstoday1999Lubkin} {}``for elucidating the quantum structure
of electroweak interactions in physics.'' In particular, they demonstrated
that the non-abelian electroweak theory could be consistently renormalized
to yield unique and precise predictions.

A key ingredient for their demonstration was the development of the
dimensional regularization technique.\citep{'tHooft:1972fi,'tHooft:1973mm,Bollini:1972ui}
That is, instead of working in precisely D=4 space-time dimensions,
they generalized the dimension to be a continuous variable so they
could compute the theory in D=4.01 or D=3.99 dimensions.

An important property of the dimensional regularization is that it
respects gauge and Lorentz symmetries.\footnote{Note, for chiral symmetries there are some subtle difficulties that
must be handled carefully. In particular, the properties of the parity
operator are dependent on the dimensionality of space-time.%
} 
This is in contrast to the other regularization schemes (e.g. cutoff
schemes, etc.) which violate these symmetries. The symmetries of the
electroweak theory play a critical role in determining the dynamics
of the particles and their interactions. Because it respects these
symmetries, dimensional regularization has become an essential tool
for the calculation of field theories.

While dimensional regularization is a powerful and elegant technique,
most examples and applications of dimensional regularization are in
the context of complex higher-order Quantum Field Theory (QFT) calculations
involving gauge and Lorentz symmetries. However, the virtues of dimensional
regularization can be exhibited without the {}``distractions'' of
the associated QFT complexities.

In the present paper, we will apply the dimensional regularization
method to a problem from an elementary undergraduate physics course,
namely the electric potential of an infinite line of charge.\citep{kaufman1969,Hans:1982vy}
The example is simple enough for the undergraduate to understand,
yet contains many of the concepts we encounter in a true QFT calculation.
We will contrast the symmetry-preserving dimensional regularization
approach with a symmetry-violating cutoff approach.

Imagining a variable number of dimensions can be a productive exercise.
To explain the weak nature of the gravitational force physicists have
recently posited the existence of {}``Extra Dimensions.'' Having
considered space-time dimensions in the neighborhood of $D=4$, we
briefly contemplate wider excursions of $D=4,5,6,...$ dimensions.

\section{Dimension Analysis: The Pythagorean Theorem}

\begin{figure}
\begin{centering}
\includegraphics[width=0.35\textwidth]{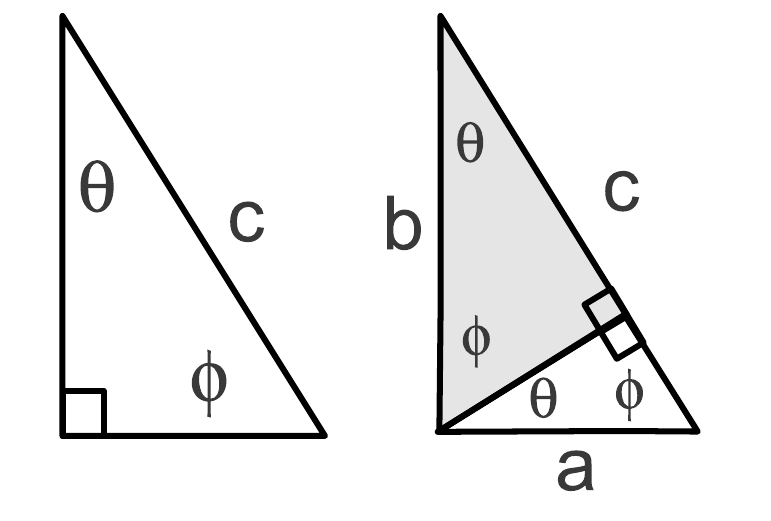} 
\par\end{centering}

\caption{a) A right triangle specified by angles $\{\theta,\phi\}$ and hypotenuse
$c$. b) The same triangular area can be described by two similar
triangles of hypotenuse $a$ and $b$.\label{fig:one}}

\end{figure}

To illustrate utility of dimensional regularization and dimensional
analysis, we warm-up with a pre-example. Our goal will be to demonstrate
the Pythagorean Theorem, and our method will be dimensional analysis.

We consider the right triangle displayed in Fig.~\ref{fig:one}-a).
From the Angle-Side-Angle (ASA) theorem, this can be uniquely specified
using the two angles $\{\theta,\phi\}$ and the hypotenuse $c$. We
now construct a formula for the area of the triangle, $A_{c}$, using
only these variables: $\{c,\theta,\phi\}$. Note that $c$ has dimensions
of length, and $\{\theta,\phi\}$ are dimensionless. From dimensional
analysis, the area of the triangle must have dimensions of length
squared. As $c$ is the only dimensional quantity, the formula for
$A_{c}$ must be of the form:

\begin{equation}
A_{c}=c^{2}f(\theta,\phi)\label{eq:areaC}\end{equation}
 where $f(\theta,\phi)$ is an unknown \emph{dimensionless} function.
Note that $f(\theta,\phi)$ cannot depend on the length $c$ as this
would spoil the dimensionless nature of $f(\theta,\phi)$.

We now observe that we can divide the original triangle of Fig.~\ref{fig:one}-a)
into two similar triangles of hypotenuse $a$ and $b$ as displayed
in Fig.~\ref{fig:one}-b). Again, using the ASA theorem, we can represent
the area of these triangles, $A_{a}$ and $A_{b}$, in terms of the
variables $\{a,\theta,\phi\}$ and $\{b,\theta,\phi\}$, respectively.
Again from dimensional considerations, these areas must be proportional
to $a^{2}$ and $b^{2}$. Thus, we obtain:

\begin{equation}
A_{a}+A_{b}=a^{2}f(\theta,\phi)+b^{2}f(\theta,\phi)\quad.\label{eq:areaAB}\end{equation}
 Because all three triangles are similar, their areas are described
by the same $f(\theta,\phi)$. It is important to note that the function
$f(\theta,\phi)$ is \emph{universal}, dimensionless, and scale-invariant.

Finally, we use {}``conservation of area'' to obtain our result.
Specifically, since the area of the original triangle $A_{c}$ is
equal to the sum of the combined $A_{a}$ and $A_{b}$, \begin{eqnarray}
A_{a}+A_{b} & = & A_{c}\quad.\label{eq:areaABC}\end{eqnarray}
 We can substitute Eqs.~(\ref{eq:areaC}) and~(\ref{eq:areaAB})
to obtain our desired result:

\begin{eqnarray}
a^{2}f(\theta,\phi)+b^{2}f(\theta,\phi) & = & c^{2}f(\theta,\phi)\nonumber \\
a^{2}+b^{2} & = & c^{2}\quad.\label{eq:pyth}\end{eqnarray}

The last equation is, of course, the Pythagorean Theorem. Clearly,
there are much simpler methods to prove this theorem; however, this
method does illustrate the power of the dimensional analysis 
approach.\footnote{In Sec.~\ref{sec:DimReg} we will use dimensional analysis to demonstrate
that we \emph{must} introduce an auxiliary scale $\mu$ in addition
to the regulator $\epsilon$. For other interesting applications of
scaling and dimensional analysis \emph{cf}. Refs.~\citep{Migdal:1977bq,physicstoday1998vogel,vogel,BernsteinFriedman}.%
} 
Additionally, we gain a new perspective on the Pythagorean Theorem
in this proof as it is linked to conservation of area.

There are instances, such as renormalizable field theory, where dimensional
analysis tools are essential to making certain calculations tractable.
The following example will illustrate some of these features.

\section{An Infinite Line of Charge}

\subsection{Statement of the Problem}

\begin{figure}
\begin{centering}
\includegraphics[width=0.45\textwidth]{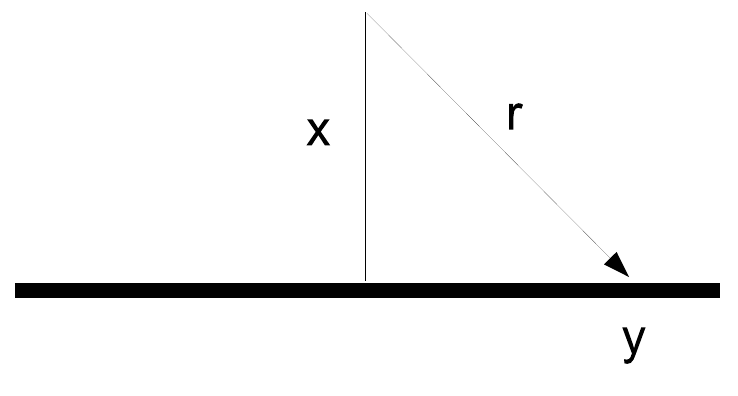} 
\par\end{centering}

\caption{Coordinate system for an infinite line of charge running in the $y$-direction
with linear charge density $\lambda=dQ/dy$. We compute the potential
$V(x)$ at a fixed perpendicular distance $x$ from the line of charge.
The distance to the element of charge $dQ$ is $r=\sqrt{x^{2}+y^{2}}$.\label{fig:two}}

\end{figure}

For our next example we consider the calculation of the electric potential
$V$ for the case of an infinite line of charge with constant linear charge
density $\lambda=dQ/dy$. The contribution to the electric potential
from an infinitesimal charge $dQ$ is given by:\footnote{We will use MKS units here so that our results reduce to the usual
undergraduate textbook expressions. %
}

\begin{equation}
dV=\,\frac{1}{4\pi\epsilon_{0}}\,\frac{dQ}{r}\quad.\label{eq:}\end{equation}
 We choose our coordinate system (cf. Fig.~\ref{fig:two}) such
that $x$ specifies the perpendicular distance from the wire, $y$
is the coordinate along the wire, and $r=\sqrt{x^{2}+y^{2}}$. Given
$\lambda=dQ/dy$ we have $dQ=\lambda dy$ and can integrate along
the length of the wire to obtain:

\begin{eqnarray}
V(x) & = & \frac{\lambda}{4\pi\epsilon_{0}}\int_{-\infty}^{+\infty}\frac{dy}{\sqrt{x^{2}+y^{2}}}=\infty\quad.\label{eq:v}\end{eqnarray}
 Unfortunately, this integral is logarithmically divergent and we
obtain an infinite result.

\subsection{Scale invariance:\label{sub:Scale-invariance:}}

If we take a closer look at this integral, we will demonstrate that
it is scale invariant. That is, if we rescale the argument $x$ by
a constant factor $k \; (x\to k\, x)$, the result is invariant.

\begin{eqnarray}
V(k\, x) & = & \frac{\lambda}{4\pi\epsilon_{0}}\int_{-\infty}^{+\infty}dy\ \,\frac{1}{\sqrt{(k\, x)^{2}+y^{2}}}\\
 & = & \frac{\lambda}{4\pi\epsilon_{0}}\int_{-\infty}^{+\infty}d(y/k)\,\frac{1}{\sqrt{x^{2}+(y/k)^{2}}}\\
 & = & \frac{\lambda}{4\pi\epsilon_{0}}\int_{-\infty}^{+\infty}dz\,\frac{1}{\sqrt{x^{2}+z^{2}}}\\
 & = & V(x)\quad.\end{eqnarray}
 In the above we have implemented the rescaling $z=y/k$. Since both
$y$ and $z$ are dummy variables and the integration limits are infinite,
the integral is unchanged. A consequence of this scale invariance
is: \begin{eqnarray}
V(x_{1}) & = & V(x_{2})\quad.\label{eq:scaling}\end{eqnarray}

At first glance, this result appears to be a disaster since the usual
purpose of the electric potential is to compute the work $W$ via
the formula \begin{equation}
W/Q=\Delta V=V(x_{2})-V(x_{1})\end{equation}
 or to compute the electric field via

\begin{equation}
\vec{E}=-\vec{\nabla}V\quad.\end{equation}
 As Eq.~(\ref{eq:scaling}) suggests $V(x_{2})-V(x_{1})=0$, this
implies that our attempts to compute the work $W$ or the electric
field $\vec{E}$ will be meaningless.

We now understand why it is fortunate that $V(x)$ is infinite as
infinite numbers have some unusual properties. For example, given
a finite constant $c$ we can write (schematically) $\infty+c=\infty$
which implies $\infty-\infty=c$. We now understand that even though
we have $V(x_{1})=V(x_{2})$, because these quantities are infinite
we can still find that the difference is non-zero: $\delta V=V(x_{2})-V(x_{1})\not=0$.
The challenge is that the difference of two infinite quantities is
ambiguous. That is, how can we tell if $\infty-\infty=c_{1}$ or $\infty-\infty=c_{2}$
is the correct physical result?

The solution is that we must regularize the infinite quantities so
that we can uniquely extract the difference.

\section{Cutoff Regularization:}

\subsection{Cutoff Regularization Computation}

We will first regularize the integral using a simple cutoff method.
That is, instead of considering an infinite wire, we will compute
the potential for a finite wire of length $2L$. In this instance,
the potential becomes:\footnote{For simplicity, we will calculate the potential at the mid-point of
the wire; the general case is more complicated algebraically, but
yields the same result in the $L\to\infty$ limit. %
}

\begin{eqnarray}
V(x) & = & \frac{\lambda}{4\pi\epsilon_{0}}\int_{-L}^{+L}dy\frac{1}{\sqrt{x^{2}+y^{2}}}\nonumber \\
 & = & \frac{\lambda}{4\pi\epsilon_{0}}\Log\left[\frac{+L+\sqrt{L^{2}+x^{2}}}{-L+\sqrt{L^{2}+x^{2}}}\right]\quad.\label{eq:cutoff}\end{eqnarray}
 We make the following observations. 
\begin{itemize}
\item The result is finite. 
\item In addition to the physical length scale $x$, $V(x)$~depends on
an artificial regulator $L$. 
\item We cannot remove the regulator $L$ without $V(x)$~becoming singular. 
\item The result for $V(x)$ violates a symmetry of the original problem---translation
invariance. 
\end{itemize}

\subsection{Computation of $E$ and $\delta V$}

Even though $V(x)$ depends on the artificial regulator $L$, we observe
that all physical quantities are independent of this regulator in
the limit $L\to\infty$. Specifically, for the electric field we have:

\begin{eqnarray}
E(x) & = & \frac{-\partial V(x)}{\partial x}=\frac{\lambda}{2\pi\epsilon_{0}x}\frac{L}{\sqrt{L^{2}+x^{2}}}\nonumber \\
 &  & \ {\longrightarrow\atop {\scriptstyle {L\to\infty}}}\ \frac{\lambda}{2\pi\epsilon_{0}x}\end{eqnarray}
 and for the potential difference (proportional to the electric work
$W$) we have:

\-\begin{eqnarray}
\delta V & = & V(x_{1})-V(x_{2}){\longrightarrow\atop {\scriptstyle {L\to\infty}}}\ \frac{\lambda}{4\pi\epsilon_{0}}\Log\left[\frac{x_{2}^{2}}{x_{1}^{2}}\right]\quad.\label{eq:deltaV}\end{eqnarray}
 As we observed in Sec.~\ref{sub:Scale-invariance:}, $\delta V$
is finite even though it is the difference of two infinite terms $V(x_{1})$
and $V(x_{2})$. The regulator $L$ allows us  unambiguously to extract
the finite difference $\delta V$, at which point the regulator can
be discarded ($L\to\infty$). The fact that the physical quantities
$E(x)$ and $\delta V$ are independent of the unphysical regulator
is a essential property of any regularization method. We will discuss
this further in Sec.~\ref{sec:The-Renormalization-Group}.

\subsection{Broken Translational Symmetry: }

Notice that the presence of the cutoff $L$ breaks the translation
symmetry of the original problem. That is, for a truly infinite wire,
our position in the $y$-direction is inconsequential; however, for
a finite wire this is no longer the case. Specifically, if we shift
our $y$-position by a constant $c$ to $y\to y'=y+c$, our result
becomes:

\begin{eqnarray}
V(x) & = & \frac{\lambda}{4\pi\epsilon_{0}}\int_{-L+c}^{+L+c}dy\frac{1}{\sqrt{x^{2}+y^{2}}}\nonumber \\
 & = & \frac{\lambda}{4\pi\epsilon_{0}}\Log\left[\frac{+(L+c)+\sqrt{(L+c)^{2}+x^{2}}}{-(L-c)+\sqrt{(L-c)^{2}+x^{2}}}\right]\quad.\label{eq:cutoff2}
\nonumber\\
\end{eqnarray}
 Clearly we have lost the translation invariance $y\to y'=y+c$.

While preserving symmetries is not of paramount importance in this
simple example, it is essential for certain field theory calculations.
We now repeat this calculation, but instead using dimensional regularization
which will preserve the translational symmetry.

\subsection{Recap}

In summary, we find that our problem is solved at the expense of 1)
an extra scale $L$ which serves both to regulate the infinities and
provide an auxiliary length scale, and 2) a broken symmetry---translational
invariance.

\section{Dimensional Regularization \label{sec:DimReg}}

\def\strut{\rule[-8pt]{0pt}{20pt}}
\begin{table}
\begin{tabular}{|c||c|c|c|c|c|c|}
\hline \strut
$n$  & $\Omega_{n}$  & $\Gamma(n/2)$  & Object  & $V_{n}$  & Surface  & $S_{n-1}$\tabularnewline
\hline
\hline \strut
1  & $2$  & $\sqrt{\pi}$  & Line  & $2R$  & Point  & $2$\tabularnewline
\hline  \strut
2  & $2\pi$  & 1  & Disk  & $\pi R^{2}$  & Line  & $2\pi R$\tabularnewline
\hline  \strut
3  & $4\pi$  & $\frac{1}{2}\,\sqrt{\pi}$  & 3-Ball  & $\frac{4\pi}{3}R^{3}$  & 2-Sphere  & $4\pi R^{2}$\tabularnewline
\hline  \strut
4  & $2\pi^{2}$  & 1  & 4-Ball  & $\frac{\pi^{2}}{2}R^{4}$  & 3-Sphere  & $2\pi^{2}R^{3}$\tabularnewline
\hline  \strut
5  & $\frac{8\pi^{2}}{3}$  & $\frac{3}{4}\,\sqrt{\pi}$  & 5-Ball  & $\frac{8\,\pi^{2}}{15}R^{5}$  & 4-Sphere  & $\frac{8\pi^{2}}{3}R^{4}$\tabularnewline
\hline
\end{tabular}
\caption{Angular integration measure $\Omega_{n}$ as a function of dimension
$n$. The surface of the $n$-dimensional volume $V_{n}$ is an $(n-1)$-dimensional
manifold $S_{n-1}$. We recognize $\Omega_{2}$ as the circumference
of the unit circle, $\Omega_{3}$ as the surface area of the unit
sphere, and $\Omega_{4}$ as the 3-surface of the 4-dimensional unit
hypersphere. See Appendix~\ref{sub:1-Dimension} for details. \label{tab:omega}}
\end{table}

\subsection{Generalization to Arbitrary Dimension }

The central idea of dimensional regularization is to compute $V(x)$
in $n$-dimensions where $n$ is not necessarily an integer.\citep{Bollini:1972ui,'tHooft:1973mm,'tHooft:1972fi}
We can generalize the integration of Eq.~(\ref{eq:v}) by replacing
the one-dimensional integration $\int dy$ by the general $n$-dimension
result. Specifically, we make the replacement:\footnote{Here, $V_{n}$ with a subscript represents volume, and $V(x)$ represents
the potential. %
}

\begin{equation}
\int_{-\infty}^{+\infty}\, dy=\int dV_{1}\longrightarrow\int dV_{n}=\int d\Omega_{n}\,\int_{0}^{+\infty}y^{n-1}dy\,.\end{equation}
 where the angular integration measure is given by

\begin{eqnarray}
\Omega_{n}=\int d\Omega_{n} & = & \frac{2\pi^{n/2}}{\Gamma\left(\frac{n}{2}\right)}\equiv\frac{n\,\pi^{n/2}}{\Gamma\left(\frac{n}{2}+1\right)}\quad.\end{eqnarray}
 Here, $\Omega_{n}$ is the solid-angle in $n$-dimensions, and we
have used $\Gamma(z+1)=z\,\Gamma(z)$ where $\Gamma$ is the Gamma
function. In Appendix~\ref{sub:1-Dimension} we provide additional
explanation, and verify that $\Omega_{n}$ yields the expected results
for integer dimensions as tabulated in Table~\ref{tab:omega}.

\subsection{Computation of V in arbitrary dimensions}

The generalized formula for $V(x)$ now reads:\citep{Hans:1982vy}

\begin{eqnarray}
V(x) & = & \frac{\lambda}{4\pi\epsilon_{0}}\int\, d\Omega_{n}\int_{0}^{+\infty}\ \frac{y^{n-1}}{\mu^{n-1}}\ \frac{dy}{\sqrt{x^{2}+y^{2}}}\,.\label{eq:Vdimreg}\end{eqnarray}
 Note that we are forced to introduce an auxiliary scale factor of
$\mu^{n-1}$, where $\mu$ has units of length, to ensure $V(x)$
has the correct dimension.\footnote{Since the factor $\lambda/(4\pi\epsilon_{0})$ has units of potential,
the integral must be dimensionless. %
} 
Replacing $n=1-2\epsilon$ to facilitate expanding about $n=1$ we
obtain
\begin{eqnarray}
V(x) & = & \frac{\lambda}{4\pi\epsilon_{0}}\frac{\Gamma\left[\frac{1-n}{2}\right]}{\left(\frac{x}{\mu}\sqrt{\pi}\right)^{1-n}}\nonumber \\
 & = & \frac{\lambda}{4\pi\epsilon_{0}}\left(\frac{\mu^{2\epsilon}}{x^{2\epsilon}}\,\frac{\Gamma[\epsilon]}{\pi^{\epsilon}}\right)\quad.\label{eq:dimreg}\end{eqnarray}

We make the following observations about the dimensionally regularized
result. 
\begin{itemize}
\item $V(x)$ depends on an artificial regulator $\epsilon$ which is dimensionless. 
\item $V(x)$ depends on an auxiliary scale $\mu$ which has dimensions
of length. 
\item If we remove either the regulator $\epsilon$ or the auxiliary scale
$\mu$ then $V(x)$ will become ill-defined. 
\item The dimensional regularization preserves the translation invariance
of the original problem. 
\end{itemize}
It is interesting to contrast this result with the cutoff regularization
method where $L$ serves as both the regulator and the auxiliary scale.

\subsection{Computation of $E$ and $\delta V$}

For the potential difference we find

\begin{eqnarray}
\delta V & = & V(x_{1})-V(x_{2}){\longrightarrow\atop \epsilon\to0}\ \frac{\lambda}{4\pi\epsilon_{0}}\Log\left[\frac{x_{2}^{2}}{x_{1}^{2}}\right]\end{eqnarray}
 and for the electric field we obtain:

\begin{eqnarray}
E & = & \frac{-\partial V(x)}{\partial x}=\frac{\lambda}{4\pi\epsilon_{0}}\left[\frac{2\epsilon\mu^{2\epsilon}\Gamma[\epsilon]}{\pi^{\epsilon}x^{1+2\epsilon}}\right]\nonumber \\
 & {\longrightarrow\atop \epsilon\to0} & \ \frac{\lambda}{2\pi\epsilon_{0}}\,\frac{1}{x}\quad.\label{eq:efield}\end{eqnarray}
 As before, we observe that all physical quantities are independent
of both the regulator $\epsilon$ and the auxiliary scale $\mu$.

\vspace{-0.3cm}

\subsection{Recap}

In conclusion we find that the problem for $V(x)$ is solved at the
expense of an artificial regulator $\epsilon$ and an auxiliary scale
$\mu$. We also note the regulator $\epsilon$ and auxiliary scale
$\mu$ are separate entities in contrast to the cutoff regularization
method where the length $L$ plays both roles. Additionally, translational
invariance symmetry is preserved. The fact that dimensional regularization
respects symmetries makes this technique indispensable for field theory
calculations involving gauge symmetries and Lorentz symmetries.

\section{Renormalization}

Having demonstrated two separate methods to regularize the infinities
that enter the calculation of $V(x)$, we now turn to renormalization.

While physical quantities such as the work $W\sim\delta V$ and the
electric field $\vec{E}\sim-\vec{\nabla}V$ are derived from $V(x)$,
the potential itself is not a physical quantity. In particular, we
can shift the potential by a constant $c$, $V\to V+c$, and the physical
quantities will be unchanged.

To illustrate this point, let's expand $V(x)$ of Eq.~(\ref{eq:dimreg})
in powers of $\epsilon$:

\begin{eqnarray}
V(x) & = & \frac{\lambda}{4\pi\epsilon_{0}}\left[\frac{1}{\epsilon}+\ln\left[\frac{e^{-\gamma_{E}}}{\pi}\right]+\ln\left[\frac{\mu^{2}}{x^{2}}\right]+{\cal O}(\epsilon)\right]\,.
\nonumber\\
\end{eqnarray}
 Here, $\gamma_{E}\simeq0.577216$ is the Euler constant which arises
from expanding the Gamma function $\Gamma[\epsilon]\sim\frac{1}{\epsilon}-\gamma_{E}$.

Let us now invent a Minimal Subtraction (MS) prescription. We have
the freedom to shift $V(x)$ by a constant, and we choose this constant
to eliminate the $1/\epsilon$ term:

\begin{eqnarray}
V_{MS}(x) & = & \frac{\lambda}{4\pi\epsilon_{0}}\left[\phantom{\frac{1}{\epsilon}+}\ln\left[\frac{e^{-\gamma_{E}}}{\pi}\right]+\ln\left[\frac{\mu^{2}}{x^{2}}\right]+{\cal O}(\epsilon)\right]\,.
\nonumber\\
\end{eqnarray}
 We can go even further and invent a Modified Minimal Subtraction
($\msbar$) prescription to eliminate the $\ln[e^{-\gamma_{E}}/\pi]$
term as well:

\begin{eqnarray}
V_{\msbar}(x) & = & \frac{\lambda}{4\pi\epsilon_{0}}\left[\phantom{\frac{1}{\epsilon}+\ln\left[\frac{e^{-\gamma_{E}}}{\pi}\right]+}\ln\left[\frac{\mu^{2}}{x^{2}}\right]+{\cal O}(\epsilon)\right]\,.
\nonumber\\ 
\end{eqnarray}
 After renormalization we can remove the regulator ($\epsilon\to0$),
but not the auxiliary scale $\mu$. Recall that without an auxiliary
scale to generate a dimensionless ratio $\mu/x$ we could not have
any substantive $x$-dependence.

In addition to the $\mu$-dependence we will also have renormalization
scheme dependence in $V(x)$. However, physical observables \emph{must}
be independent of the auxiliary scale $\mu$ and the particular renormalization
scheme. For example, the computed potential differences yield identical
results when calculated consistently in a single renormalization scheme:

\begin{eqnarray}
V_{MS}(x_{1})-V_{MS}(x_{2}) & = & \delta V=V_{\msbar}(x_{1})-V_{\msbar}(x_{2})\quad.
\nonumber\\
\end{eqnarray}
 Here, the results of the Minimal Subtraction (MS) and the Modified
Minimal Subtraction ($\msbar$) are identical for \emph{physical}
quantities.

However, if you mix renormalization schemes inconsistently you will
obtain non-sensible results that \emph{are} dependent on the choice
of scheme:\footnote{The reader is invited to verify that the computation of the electric
field $\vec{E}(x)$ in a consistent renormalization scheme yields
the previous results of Eq.~(\ref{eq:efield}), and an inconsistent
application of the schemes does not. %
}

\begin{eqnarray}
V_{\msbar}(x_{1})-V_{MS}(x_{2}) & \not= & \delta V\not=V_{MS}(x_{1})-V_{\msbar}(x_{2})\quad.
\nonumber\\
\end{eqnarray}

\subsection{Connection to QFT}

This elementary problem of the infinite line charge contains all the
key concepts of the dimensional regularization and renormalization
that we encounter in the full QFT radiative calculations. For example,
in the radiative Quantum Chromodynamics (QCD) calculation of the Drell-Yan
process ($q\bar{q}\to\gamma^{*}\to\mu^{+}\mu^{-}$) we encounter the
following infinite expression:\footnote{Cf. Ref.~\citep{potter1997}, Eq.~(46) and Eq.~(47).%
}

\begin{eqnarray}
\frac{D(\epsilon)}{\epsilon} & = & \left(\frac{4\pi\mu^{2}}{Q^{2}}\right)^{\epsilon}\,\frac{\Gamma(1-\epsilon)}{\Gamma(1-2\epsilon)}\nonumber \\
 & \sim & \frac{1}{\epsilon}-\ln\left(\frac{e^{+\gamma_{E}}}{4\pi}\right)+\ln\left(\frac{\mu^{2}}{Q^{2}}\right)\quad.\end{eqnarray}
 In this equation, $Q$ represents the characteristic energy scale. 
This is the independent variable that is analogous to $x$ in our
example. While this is for a 4-dimensional QCD calculation, the structure
of the divergent term is remarkably similar to our simple one-dimensional
example above. For the QCD calculation, the Minimal Subtraction ($MS$)
prescription for this Drell-Yan calculation eliminates the $1/\epsilon$
term, and the Modified Minimal Subtraction ($\msbar$) prescription
eliminates the $1/\epsilon-\ln[e^{+\gamma_{e}}/(4\pi)]$ so that only
the $\ln[\mu^{2}/Q^{2}]$ remains.

\section{The Renormalization Group Equation\label{sec:The-Renormalization-Group}}

\subsection{Physical Observables:}

The fact that the physical observables are independent of the unphysical
auxiliary scale $\mu$ is simply a consequence of the renormalization
group equation (RGE):\footnote{For an excellent pedagogical analysis of the renormalization group
equation \emph{cf}. Ref.\citep{Delamotte:2002vw}.%
}

\begin{eqnarray}
\mu\,\frac{d\sigma}{d\mu} & = & 0\label{eq:RGE}\end{eqnarray}
 where $\sigma$ represents any physical observable. Thus, the renormalization
group equation implies that the electric field $\vec{E}=\vec{\nabla}V$
and the work $W=\delta V$ are also independent of the $\mu$ scale:
\begin{equation}
\mu\,\frac{dE}{d\mu}=0\qquad\qquad\mu\,\frac{dW}{d\mu}=0\quad.\end{equation}
 These results are implicit in the final expression for the physical
quantities $E$ and $V$.

\vspace{-0.3cm}

\subsection{Relating Perturbative \& Non-Perturbative Functions}

While the result of Eq.~(\ref{eq:RGE}) appears to be almost trivial
in the above example, this yields a very important result when applied
to scattering processes involving non-perturbative hadronic particles
(proton, nucleons, etc.). We can write the physical cross section
$\sigma$ as a product of a non-perturbative distribution $f$ which
describes the soft (low energy) physics, and a perturbative term $\omega$
which describes the hard (high energy) physics:\footnote{More precisely, $f$ is a {}``parton distribution function,'' and
$\omega$ is a {}``hard-scattering cross section.'' The cross section
$\sigma$ is a convolution $\sigma=f\otimes\omega$ which can be decomposed
by taking Mellin moments; hence, the discussion of this section applies
formally to the Mellin transforms of $f$ and $\omega$. %
} 
\begin{equation}
\sigma=f\,\omega\quad.\end{equation}
 Differentiating with respect to $\ln\mu$ and applying the chain
rule we find

\begin{equation}
\frac{d\sigma}{d\ln\mu}=0=\frac{df}{d\ln\mu}\,\omega+f\,\frac{d\omega}{d\ln\mu}\end{equation}
 where we have used Eq.~(\ref{eq:RGE}). Rearranging terms, we place
all the $f$ dependence on the left-hand-side (LHS) and the $\omega$
dependence on the right-hand-side (RHS),

\begin{equation}
\frac{1}{f}\,\frac{df}{d\ln\mu}=-\gamma=\frac{1}{\omega}\,\frac{d\omega}{d\ln\mu}\quad.\label{eq:gamma}\end{equation}
 We introduce a separation constant\footnote{Unless $f$ and $\omega$ are trivially related, the most reasonable
solution for this type of differential equation is that both the LHS
and RHS of Eq.~(\ref{eq:gamma}) equal a separation constant, $-\gamma$. %
} 
$-\gamma$. We note the LHS of Eq.~(\ref{eq:gamma}) depends only
on the non-perturbative quantity $f$; therefore, the LHS is (in principle)
incalculable. Conversely, the RHS of Eq.~(\ref{eq:gamma}) depends
only on the perturbative quantity $\omega$. Therefore, the RHS is
calculable in perturbation theory, and we can use this to compute
$-\gamma$.

Having computed $-\gamma$, we can solve Eq.~(\ref{eq:gamma}) for
$f$ to obtain
\footnote{The term $-\gamma$ is referred to as the \emph{anomalous dimension}.
It is a \emph{dimension} because it determines the $\mu$-scaling
dimension of $f$ in Eq.~(\ref{eq:anomalousDim}). It is \emph{anomalous}
because if $f$ satisfied exact scaling, $f$ would be invariant under
a scale change ($\mu_{1}\to\mu_{2}$); so $f=\mu^{0}=const$, and
any non-zero value for $-\gamma$ would be anomalous. %
}

\begin{equation}
f\sim\mu^{-\gamma}\quad.\label{eq:anomalousDim}\end{equation}
 Equation~(\ref{eq:anomalousDim}) is a remarkable result! Even though
$f$ was an incalculable non-perturbative quantity, we are able to
find the $\mu$-dependence for this function. Thus, the renormalization
group equation has allowed us to compute the $\mu$-dependence of
an incalculable quantity by relating the (incalculable) non-perturbative
$df/f$ to the (calculable) perturbative $d\omega/\omega=-\gamma$.

\section{Extra Dimensions}

\begin{table}
\begin{tabular}{|c|c|c|c|}
\hline 
$D_{{\rm eff}}$  & $E(r)$  & $V(r)$  & Example\tabularnewline
\hline
\hline 
3  & $\frac{1}{r^{2}}$  & $\frac{1}{r}$  & Point charge\tabularnewline
\hline 
2  & $\frac{1}{r^{1}}$  & $\ln\, r$  & Line charge\tabularnewline
\hline 
1  & $\frac{1}{r^{0}}$  & $r$  & Sheet charge\tabularnewline
\hline
\end{tabular}

\caption{Example charge configurations that illustrate $D_{{\rm eff}}=\{3,2,1\}$
effective dimensions. \label{tab:deff}}

\end{table}

\subsection{E and V in arbitrary dimensions}

In the above example, we used the mathematical trick of generalizing
the number of integration dimensions from an integer to a continuous
parameter. While we only let the dimension stray by $2\epsilon$,
it is useful to consider more drastic shifts as in the case of {}``Extra-Dimensions''
which have recently been hypothesized.\citep{ArkaniHamed:1998rs,Randall:1999ee}
In this section, we provide an example of a dimensional transmutation 
where the effective dimension $D_{{\rm eff}}$ changes from
one integer to another as we probe the system at different scales.

For example, we can generalize the $r$-dependence of the potential
and electric field in for the case of $D$-dimensions as:
\footnote{Note, for the special case D=2 the potential $V(r)$ has a logarithmic
form; see Table~\ref{tab:deff} for details. %
}

\begin{equation}
V(r)\sim\frac{1}{r^{D-2}}\qquad\qquad E(r)\sim\frac{1}{r^{D-1}}\quad.\end{equation}
 A quick check will verify that this reproduces the usual expressions
in ordinary $D=3$ spacial dimensions. Additionally, in 3-dimensions
we can create charge distributions that mimic lower order spatial
dimensions. This is illustrated in Table~\ref{tab:deff}. For a (zero-dimensional)
point-charge in 3-dimensions, according to Gauss's law the electric
field lines spread out on a surface of $D-1=2$ dimensions, and we
observe $E(r)\sim1/r^{2}$. Similarly, for a (one-dimensional) line-charge,
our space is now effectively $D=2$ dimensional; hence the electric
field lines spread out on a surface of $D-1=1$ dimension, and we
observe $E(r)\sim1/r$. Finally, for a (two-dimensional) sheet-charge,
our space is now effectively $D=1$ dimensional; hence the electric
field lines spread out on in $D-1=0$ dimensions, and we observe $E(r)\sim1/r^{0}=constant$.

\begin{figure}
\begin{centering}
\includegraphics[width=0.45\textwidth]{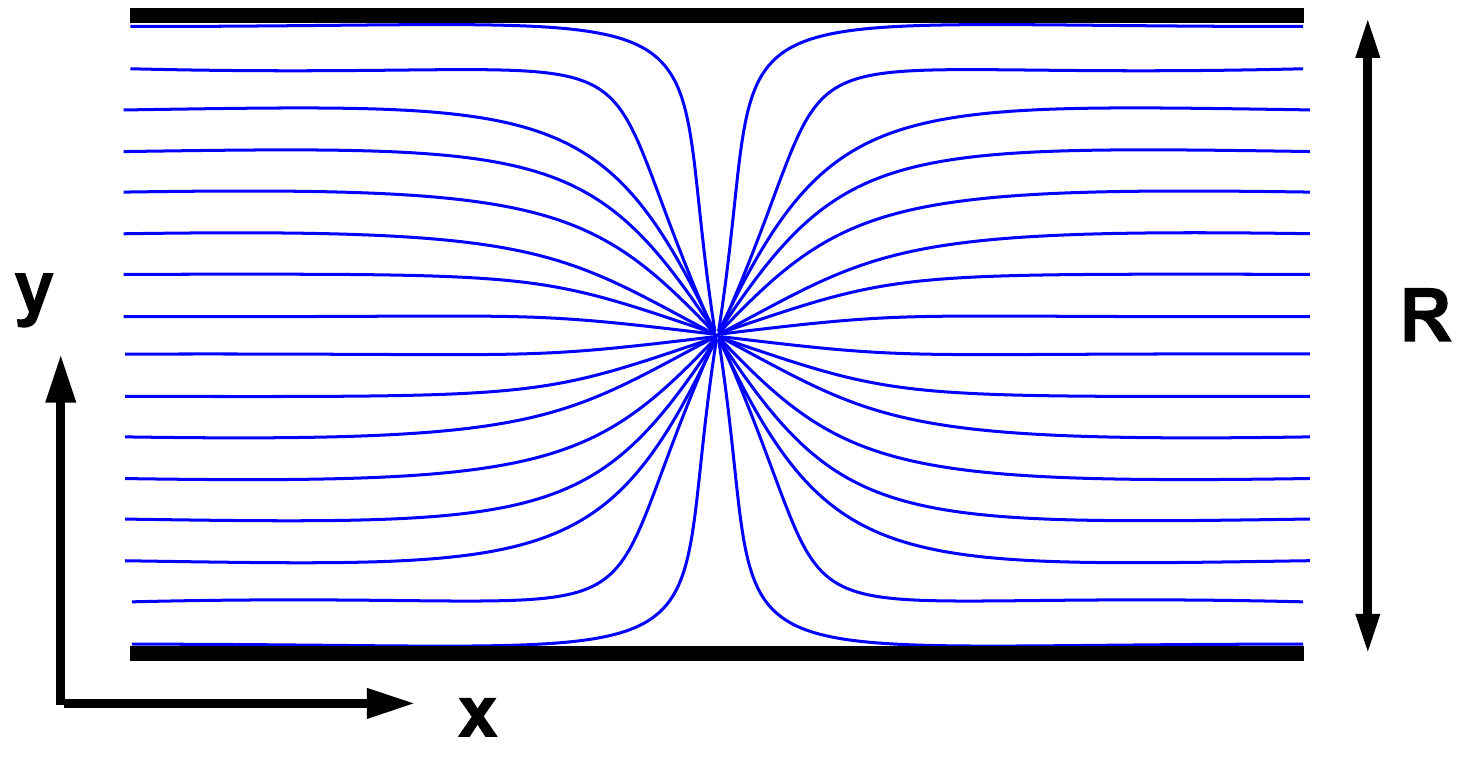} 
\par\end{centering}

\caption{Electric field for a point charge confined in one infinite dimension
$(x)$ and one finite dimension $(y)$ of scale $R$. \label{fig:efield}}

\end{figure}

Figure~\ref{fig:efield} displays the electric field lines for a
point charge confined to one infinite dimension $(x)$ and one finite
(or compact) dimension $(y)$ of scale $R$. We observe that if we
examine the electric field at scales small compared to the compact
dimension $R$ $(r\ll R)$, we find the the electric field lines spread
out in 2 dimensions and we obtain the usual 2-dimensional result $\vec{E}(r)\sim1/r$. 
Conversely, if we examine the electric field at distance scales large
compared to the compact dimension $R$ $(r\gg R)$, we find the 1-dimensional
result $\vec{E}(r)\sim constant$. In this example, the effective
dimension of our space changes as we move from small ($D=2$) to large
length scales ($D=1$).

\section{Conclusions\label{sec:conclusions}}

In this paper we have computed the potential of an infinite line of
charge using dimensional regularization. By contrasting this calculation
with the conventional cutoff approach, we demonstrated that dimensional
regularization respects the symmetries of the problem---namely, translational
invariance. The dimensional regularization requires that we introduce
a regulator $\epsilon$ and an auxiliary length scale $\mu$. We then
renormalized the potential to eliminate the $1/\epsilon$ singularities. 
This potential was finite and independent of the regulator $\epsilon$,
but it depended on the particular renormalization scheme and renormalization
scale $\mu$. However, we demonstrated that all physical observables
$(E,\delta V)$ were scheme and scale invariant.

As this example exhibits many of the key features of dimensional regularization
as applied to QFT, it provides an excellent opportunity to understand
the virtues of this regularization method without the complications
of gauge symmetries. As such, this example serves as an ideal pedagogical
study.

\section*{Acknowledgment}

This work is based on lectures presented at the CTEQ Summer Schools
on QCD Analysis and Phenomenology \texttt{(http://www.cteq.org).}
We thank Matthew Bernstein, Bryan Field, Howie Haber, Robert Jaffe,
and John Ralston for valuable discussions. We also thank the AJP reviewers
for helpful suggestions. F.I.O. acknowledges the hospitality of Argonne
National Laboratory and CERN where a portion of this work was performed.
This work is supported by the U.S. Department of Energy under grant
DE-FG02-04ER41299, the Lightner-Sams Foundation.

\section{Appendix\label{sub:1-Dimension}}

\subsection{3-Dimensions}

The volume of a 3-sphere ($V_{3}$) in spherical coordinates is a
product of the angular and radial integrals:

\begin{eqnarray}
V_{3} & = & \int d\Omega_{3}\int_{0}^{R}r^{2}dr=\int_{0}^{2\pi}d\phi\int_{0}^{\pi}\sin\theta\, d\theta\int_{0}^{R}r^{2}dr\nonumber \\
 & = & \frac{4\pi}{3}R^{3}\quad.\end{eqnarray}
 Note that the angular integral $\int\Omega_{3}$ is dimensionless
while the radial integral $\int r^{2}dr$ carries the dimensions.

For the 2-dimensional surface area ($S_{2}$), we can use the above
$V_{3}$ integral with a $\delta$-function $\delta(r-R)$ to constrain
us to the surface: \begin{equation}
S_{2}=\int d\Omega_{3}\int_{0}^{R}dr\, r^{2}\delta(r-R)=4\pi R^{2}\quad.\end{equation}

\subsection{$n$-Dimensions}

Having established the familiar 3-dimensional case, we can generalize
to $n$-dimensions:

\begin{equation}
V_{n}=\int d\Omega_{n}\,\int_{0}^{R}r^{n-1}dr=\Omega_{n}\frac{R^{n}}{n}\label{eq:vn}\end{equation}
 and the $(n-1)$-dimensional surface area ($S_{n-1}$) of the $n$-dimensional
volume $V_{n}$ is:

\begin{equation}
S_{n-1}=\int d\Omega_{n}\,\int_{0}^{R}dr\, r^{n-1}\,\delta(r-R)=\Omega_{n}\, R^{n-1}\quad.\label{eq:sn}\end{equation}
 With the above we have the general relation:

\begin{equation}
\frac{V_{n}}{S_{n-1}}=\frac{R}{n}\quad.\end{equation}
 Additionally, we find the following relation:

\begin{equation}
\frac{dV_{n}}{dR}=S_{n-1}\quad.\end{equation}
 This demonstrates that the derivative (or boundary) of the volume
is the surface area, $\partial V=S$.

\subsection{1-Dimension}

As the 1-dimensional case has a subtle factor of 2, we compute this
explicitly. Using Eq.~(\ref{eq:vn}) we find the volume of a 1-dimensional
line to be:

\begin{equation}
V_{1}=\int dV_{1}=\int d\Omega_{1}\,\int_{0}^{R}r^{0}\, dr=2R\quad.\end{equation}
 Note, this result is not $R$ but $2R$ as the 1-dimensional line
extends from $-R$ to $+R$.

In the notation of Eq.~(\ref{eq:v}) we have (with $R\to\infty$)

\begin{equation}
\int dV_{1}=\int d\Omega_{1}\,\int_{0}^{+\infty}\, dy=2\int_{0}^{+\infty}\, dy=\int_{-\infty}^{+\infty}\, dy\quad.\end{equation}
 Thus, we can make the replacement $\int_{-\infty}^{+\infty}\, dy\to\int dV_{1}$,
and the $n$-dimensional generalization is then:

\begin{equation}
\int_{-\infty}^{+\infty}\, dy=\int dV_{1}\longrightarrow\int dV_{n}=\int d\Omega_{n}\,\int_{0}^{+\infty}y^{n-1}dy\,.\end{equation}
 Eq.~(\ref{eq:v}) for the potential $V(x)$ then becomes:

\begin{eqnarray}
V(x) & = & \frac{\lambda}{4\pi\epsilon_{0}}\int d\Omega_{n}\,\int_{0}^{+\infty}y^{n-1}\frac{dy}{\sqrt{x^{2}+y^{2}}}\quad.\label{eq:v2}\end{eqnarray}
 Note that Eq.~(\ref{eq:v2}) is \emph{not} dimensionally correct
as the factor $y^{n-1}$ will need to be compensated by introducing
an auxiliary scale factor as we do in Eq.~(\ref{eq:Vdimreg}).

\bibliographystyle{plunsrt}
\bibliography{dimreg}
\end{document}